%% file: main_arxiv.tex
\documentclass[12pt]{article}
\pdfoutput=1
\usepackage{natbib}
 \bibpunct{(}{)}{;}{a}{,}{,}
\usepackage{graphicx}
\usepackage{amsmath,amssymb,amsthm}
\usepackage{bm}
\usepackage{rotating}
\usepackage[lined,ruled]{algorithm2e}
\usepackage{multirow}
\usepackage{rotating}
\usepackage{multicol}
\usepackage{titlesec}
\usepackage{latexsym}
\usepackage[pdftex,colorlinks=true,linkcolor=blue,citecolor=blue,urlcolor=blue,bookmarks=false,pdfpagemode=None]{hyperref}
\usepackage{url}
\usepackage{verbatim}
\usepackage{fancyhdr}
\usepackage{setspace}
\usepackage{paralist}
\usepackage{lineno}
\usepackage[top=1in, bottom=1in, left=1in, right=1in]{geometry}
\usepackage{lscape}
\usepackage{dcolumn}

\newcolumntype{.}{D{.}{.}{-1}}

\newtheorem{theorem}{Theorem}
\newtheorem{proposition}[theorem]{Proposition}
\newtheorem*{remark}{Remark}

\newcommand{\MLE}{{\rm MLE}}

\newcommand{\R}{{\rm R}}

\renewcommand{\bar}{\overline}

\newcommand{\bigO}[1]{\ensuremath{\mathop{}\mathopen{}\mathcal{O}\mathopen{}\left(#1\right)}}

\newcommand{\Ex}{\mathbb{E}}
\renewcommand{\Re}{\mathbb{R}}

\newcommand{\SNR}{\mathrm{SNR}}

\newcommand{\inv}{{-1}}

\begin{document}
\pagestyle{empty}

\title{
Analyzing statistical and computational tradeoffs\\ of estimation procedures
% Analyzing statistical and computational tradeoffs\\ to optimize inference on large data sets
\protect\thanks{Daniel L. Sussman is a Postdoctoral Fellow in the Department of Statistics at Harvard University (\href{mailto:daniellsussman@fas.harvard.edu}{daniellsussman@fas.harvard.edu}). Alexander Volfovsky is an National Science Foundation Mathematical Sciences Postdoctoral Research Fellow in the Department of Statistics at Harvard University (\href{mailto:volfovsky@fas.harvard.edu}{volfovsky@fas.harvard.edu}). Edoardo M.~Airoldi is an Associate Professor of Statistics at Harvard University (\href{mailto:airoldi@fas.harvard.edu}{airoldi@fas.harvard.edu}).
This work was partially supported 
 by the National Science Foundation under grants 
  CAREER IIS-1149662, TWC-1237235, IIS-1409177, and DMS-1402235,
 by the Army Research Office grant 
  MURI W911NF-11-1-0036,
 and by the Office of Naval Research under grant 
  YIP N00014-14-1-0485. 
 Edoardo M.~Airoldi is an Alfred P. Sloan Research Fellow, and a Shutzer Fellow at the Radcliffe Institute for Advanced Studies.
% Alexander Volfovsky is a National Science Foundation Mathematical Sciences Postdoctoral Research Fellow. 
%
The authors are grateful to Michael I. Jordan and Donald B. Rubin for their insightful and constructive discussion that helped to greatly improve the framing of the research.}}

\author{
 Daniel L. Sussman, Alexander Volfovsky, Edoardo M. Airoldi\\
 Department of Statistics\\
 Harvard University, Cambridge, MA 02138, USA}
\date{}

\maketitle

\newpage
\begin{abstract}
The recent explosion in the amount and dimensionality of data has exacerbated the need of trading off computational and statistical efficiency carefully, so that inference is both tractable and meaningful.
We propose a framework that provides an explicit opportunity for practitioners to specify how much statistical risk they are willing to accept for a given computational cost, and leads to a theoretical risk-computation frontier for any given inference problem.
We illustrate the tradeoff between risk and computation and illustrate the frontier in three distinct settings.
First, we derive analytic forms for the risk of estimating parameters in the classical setting of estimating the mean and variance for normally distributed data and for the more general setting of parameters of an exponential family.
The second example concentrates on computationally constrained Hodges-Lehmann estimators.
We conclude with an evaluation of risk associated with early termination of iterative matrix inversion algorithms in the context of linear regression.
% The recent explosion in the quantity and dimension of data has brought up many
% new computational and statistical challenges that must be addressed 
% simultaneously such that inference is both tractable and meaningful. We propose 
% a framework that provides an explicit opportunity for the practitioner to 
% specify how much statistical risk they are willing to accept for a given 
% computational cost. We demonstrate the strength of this framework in three
% distinct settings. First, we derive analytic forms for the risk of estimating
% parameters in general exponential families under a computational constraint
% on the calculation of sufficient statistics. The second example concentrates
% on computationally constrained Hodges-Lehmann estimators. We conclude
% with an evaluation of risk associated with early termination of
% iterative matrix inversion algorithms in the context of linear regression. 
% \keywords{computation \and risk \and exponential families}

\vfill
\end{abstract}

\newpage
\singlespacing
% \setcounter{tocdepth}{4}
% 1 Section
% 2 Subsection
% 3 Subsubsection
% 4 Paragraph
% 5 Subparagraph
%\footnotesize
\small
\tableofcontents
\normalsize
\doublespacing

\newpage
\pagestyle{plain}
\setcounter{page}{1}

\input{intro.tex}

\input{framework}

\input{normal_example.tex}

\input{asymptot.tex}
\input{hl_example.tex}
\input{matrix.tex}
\input{conc.tex}

%\begin{acknowledgements}
%\end{acknowledgements}

\bibliographystyle{spbasic}      % basic style, author-year citations
% \bibliographystyle{spmpsci}      % mathematics and physical sciences
\bibliography{biblio}

% % Non-BibTeX users please use
% \begin{thebibliography}{}
% %
% % and use \bibitem to create references. Consult the Instructions
% % for authors for reference list style.
% % Format for Journal Reference
% Author, Article title, Journal, Volume, page numbers (year)
% Author, Book title, page numbers. Publisher, place (year)
% \end{thebibliography}

\end{document}

%% file: intro.tex
% !TEX root = main.tex

% * Introduction
%     * When you have lots of data, there is a tradeoff between
%     computation costs and statistical risk/efficiency.
%     * A high level overview of this tradeoff is given by the figure
%      and an objective function that describes the tradeoff. Compare/contrast 
%      with existing discussions which concentrate on one of the two parts of the tradeoff.
%     * Introduce realistic problem settings where this tradeoff occurs:
%     no sufficient statistics, hard to compute sufficient statistics.
%     * The simplest nontrivial example is the single-look collection of algorithms: there 
%     are sufficient statistics but only one can be computed at a time. (Example: 
%     image processing). What's hard, what we find, what's surprising about it.
%     
% This dichotomy is best illustrated by the ubiquitous use
% of the ``solve'' command in \textbf{\textsf{R}} for 
% inverting matrices. 

\section{Introduction}\label{sec:intro}
The advent of massive datasets in applied fields has been 
heralded as a new age for statistics 
but these datasets are both a blessing and a curse.
They offer the opportunity to improve the precision and
efficiency of statistical methods but frequently 
these improvements come at high computational costs.
Up until recently, the computational aspects of statistics had
been largely ignored by the statistics literature with those concerns
relegated to other fields.
Statisticians have begun to explore computationally more efficient 
techniques for disparate problems using ideas like variational methods \citep{wainwright2008graphical},
parallel MCMC \citep{scott2013bayes}, stochastic gradient descent \citep{langford2009sparse,leon2012large,alekh2014reliable,toulis2014implicit}, and convex relaxations \citep{chandra2013comp}.

% This latter property had up until recently been largely ignored by the
% statistics literature
%  as tractability of a given procedure
% was dictated by the dimensionality of the problem rather than
% by the number of observations. 
Recent efforts have begun to evaluate
the computational cost of many statistical procedures and conversely the 
statistical risk of computationally efficient procedures while
introducing new procedures that balance these ideas.
\citet{kleiner2014scalable} describes a procedure known as the 
``bag of little bootstraps,'' a data splitting technique 
that allows for computationally tractable implementation of the 
bootstrap for massive datasets. \citet{wang2014statistical} study
the classical problem of covariance estimation under computational 
constraints. \citet{yang2015computational} find conditions
that ensure a Markov chain Monte Carlo sampler for Bayesian linear regression
in high dimensions will be both consistent and have fast mixing time.
 \citet{chandra2013comp} consider ``algorithm weakening''
to describe an ordering of algorithms in terms of compute time
and statistical properties. \citet{inbal2015detection} study the tradeoff
between statistical detection and computation in an edget detection framework.

A common approach is to describe compute cost in terms of algorithmic
complexity \citep{berthet2013computational,shender2013computation,bresler2014hardness,montanari2014computational}. 
% Algorithmic complexity
% describes the order of the worst case scenario run time
% as a function of the data size $n$, the data dimension
% $d$ and other components of the model and algorithm. 
Within this
framework, two algorithms are equivalent if they have the 
same worst case complexity, even if one of them is likely
to never perform in the worst case regime. While computational 
complexity is a key aspect of understanding and choosing between
algorithms, when a finer analysis is possible, it is preferred. 
\cite{chandra2013comp} provides an example of this finer analyis 
via the ``time-data tradeoff'' and our approach is in a similar spirit.
Specifically, we consider
a practical framework for addressing the tradeoff between
computational cost and statistical risk. 
This is in contrast to other works such as \cite{montanari2014computational},
who shows that some statistical procedures can be modified to provide fast algorithms 
with well understood computational gains but without assessing 
the degradation in statistical risk.

% We also seek to return to traditional statistical estimation problems
% in order to attain a deeper understanding of these tradeoffs.
% Throughout this article we will consider the problem of 
% estimating the parameter $\theta$ using data $x$ in the probability
% model $p(x|\theta)$. Unlike in the classical statistical setting,
% we consider estimators $\hat\theta$ to be not simply functions of the data
%  but also to have implemented algorithms that evaluate 
% the function. 
% In this settings, estimators can be ordered based on either their
% statistical risk, $R(\hat\theta,\theta)$, or based on 
% the minimal computational cost, $c(\hat{\theta})$, they require. 
% As such, for a fixed parameter $\theta$, sample size $n$, and a collection of estimators,
% we can ask which among the estimators minimize the 
% risk for a fixed computational cost, or equivalently those that
% minimize the computational cost for a fixed risk. Taken as a function 
% of the pair $(\theta,n)$, these estimators define the statistical
% risk and computational cost tradeoff frontier. 

Our focus for this paper will be on classical statistcal problems including
estimating the mean and variance for a normal popualation, exponential families,
robustness, and regression.
In Section~\ref{sec:framework}, of this paper we outline the basic framework 
associated with the statistical risk and computational cost tradeoff frontier.
Section~\ref{sec:norm} illustrates this framework in the normal population setting
and in Section~\ref{sec:ef} we extend these ideas to general exponential families.
We briefly consider robust estimates such as the Hodges-Lehmann example in Section~\ref{sec:hl}.
Finally, we consider extending these ideas to iterative methods for matrix inversion in Section~\ref{sec:mat}.

%% file: framework.tex
\section{Analyzing statistical and computational tradeoffs}
\label{sec:framework}

Consider the problem of estimating $\theta$ given an i.i.d. sample from the distribution $f_{\theta}$.
We suppose that the parameter is vector valued with $\theta\in \Theta\subset \Re^p$, the model is $\{f_{\theta}: \theta\in \Theta\}$ and we denote our random sample as $X_1,\dotsc,X_n\stackrel{iid}\sim f_{\theta}$ where $X_i$ is $\mathcal{X}$ valued for $\mathcal{X}\subset\Re^d$.
The estimate is denoted is $\hat{\theta}=\hat{\theta}(X_1,\dotsc,X_n)$ and the loss associated with the estimate is $\ell(\hat{\theta},\theta)$.
In classical statistical estimation theory, one seeks to in some way minimize the risk, the expected loss $R(\hat{\theta},\theta)=\Ex_{\theta}[\ell(\hat{\theta},\theta)]$, be it in terms of minimax optimality, Bayesian optimality, or other principles such as unbiasedness or equivariance \citep{lehmann1998tpe}.

We will maintain the goal of achieving a small risk but we will add the goal of computing the estimate quickly.
Formally, one need not have an algorithm to compute the estimator $\hat{\theta}:\mathcal{X}^n \mapsto \Theta$ for all values in $\mathcal{X}^n$ in order to analyze the risk and statistical properties of the estimate.
However, in our setting we will assume that each estimator comes equipped with an algorithm to compute the function of the data.
Hence, an estimate, together with its algorithm, will have a compute time $C(\hat{\theta})\in \Re^+$ that denotes the runtime of that algorithm on the data $X_1,\dotsc, X_n$. 
Altogether we now have two quantities, the risk $R(\hat{\theta},\theta)$ and the expected compute time $\Ex_\theta[C(\hat{\theta})]$.

\begin{remark}
To keep things more straightforward, the first few examples considered in this manuscript will have the property that $C(\hat{\theta})$ does not depend on the data $X_1,\dotsc,X_n$ so that the expected compute time does not depend on the parameter.
Clearly many algorithms do not fit this mold and much of the ideas we discuss apply outside the fixed computational cost setting.
\end{remark}

\begin{figure}[t!]
	\centering
	\includegraphics[width=.75\textwidth]{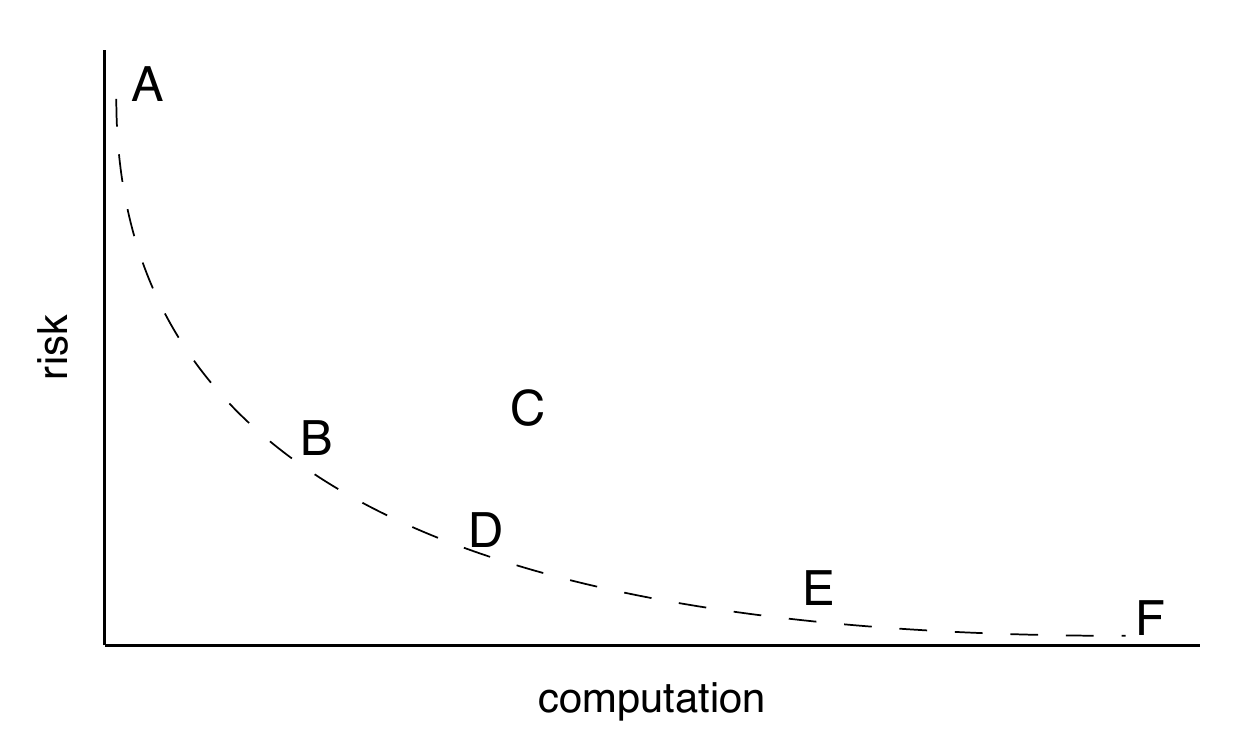}
	\caption{An illustration the risk and computation time trade-off associated with a collection of 6 estimators. 
	Estimator A is the fastest estimators but has much higher risk than the other estimators while estimator F has the lowest risk but suffers from being quite slow.
	Estimators B or D might be a good choice for a practitioner seeking to achieve a balance between time and accuracy. 
	Note that estimator C can be disregarded since both B and D are strictly better than C in terms of both risk and computation time. The dashed line depicts the theoretical risk-computation frontier.}
	%% R CODE TO MAKE THIS PLOT
	% df <- data.frame(computation=c(1,2,3.25,3,5,7.0),label=c("A","B","C","D","E","F"))
	% df$risk <- 1/df$computation
	% df$risk[3] <- df$risk[3]+.3
	% df$risk <- df$risk*(1+runif(6)*.3)
	% ggplot(df,aes(x=computation, y=risk,label=label))+geom_text()+theme_classic()+scale_x_continuous(breaks=c())+scale_y_continuous(breaks=c())+geom_line(mapping=aes(y=.85/computation),data=data.frame(computation=seq(0.75,7.5,.01),label=""),linetype=2,color="gray")
	% ggsave("~/Dropbox/Manuscript/ComputationDan/STCO_latex/figures/risk_comp_tradeoff_illustration.pdf",width=5,height=3)
	\label{fig:tradeoff}
\end{figure}

Now, consider a practitioner confronted with a collection of estimators, each with an associated algorithm. 
We denote this collection by $\{\hat{\theta}_s\}_{s\in S}$ where $S$ is some index set. 
The collection of estimates may be determined by questions such as:
% \begin{itemize}\setlength\itemsep{0em}
% \item 
How much storage is available?
% \item 
Can all the data be kept in memory or only a subset?
% \item 
How much processing power is available?
% \item 
Are there parallel or distributed systems that can be exploited?
% \end{itemize}

Each of these questions will put different constraints on the collection of estimators that delineates what is possible and ready for use.
Among the feasible estimators, the practitioner must chose an estimator $\hat{\theta}_s$ to use on the data at hand.
If the practitioner knows $R(\hat{\theta}_s, \theta)$ and $C(\hat{\theta}_s)$ for each estimate and parameter value than they can use this to make an informed decision about which estimator to choose that balances risk and computational cost.
For a given parameter value $\theta$ we can plot the risk and computation time associated with each estimator as was done in Figure~\ref{fig:tradeoff}.
This figure illustrates the idea of a computational-statistical trade-off with some estimates achieving very low risk, others being very fast, and some providing a balance between the two.
In the next section we will examine this framework in the setting estimating the mean and variance for a sample from a normal distribution before investigating exponential families in general in Section~\ref{sec:ef}.

%% file: normal_example.tex
% !TEX root = main.tex

\section{Normal example}\label{sec:norm}
To concretely illustrate the computation-statistical tradeoffs we 
will consider the simple example of estimating the population mean
and variance from a sample of independent and identically
distributed normal random variables.
We consider two computational constraints
that define the collections of algorithms that are 
available for estimation.
The first setting explores a singly indexed set of estimators for a near zero resource streaming setting. 
% The first setting we explore has no local storage of
% data which limits the collection of algorithms that can 
% accommodate a continuous stream of data where each data point
% can only be used once.
% While our 
% example for the normal model is artificial, this type of
% limitation is common when each data point is massive such
% as in image processing (REF).
The second setting generalizes the first by allowing various ways to divide the data between different aspects of the estimation.

\subsection{Standard inference}

Suppose that we observe $X_1,\dotsc,X_n \stackrel{iid}{\sim}\mathcal{N}(\mu, \sigma^2)$ and we want to estimate $\theta=(\mu,\sigma^2)$ with our loss being square error loss, $\ell(\hat{\theta},\theta)=\|\hat{\theta}-\theta\|_2^2=(\hat{\mu}-\mu)^2+(\hat{\sigma}^2-\sigma^2)^2$.
Our analysis allows for loss functions that are other linear combinations of the risks for $\mu$ and $\sigma^2$ however for ease of illustration we focus on this loss function.

Before delving into various computationally constrained estimates, consider the standard maximum likelihood
estimates (MLE) of the mean and variance for a normal. 
\begin{equation}
    \hat\mu_\MLE=\bar X=\frac{1}{n}\sum_{i=1}^n X_i \quad \text{and}\quad 
\hat\sigma^2_\MLE=\bar{X^2}-\bar X^2= \frac{1}{n}\sum_{i=1}^n X_i^2 - \bar{X}^2.\label{eq:mle}
\end{equation}
To compute this estimate 
we require two operations or {\em looks} at each data point to compute the sufficient statistics $\bar X$
and $\bar{X^2}$---that is we need to temporarily store each data point in order 
to perform a local operation before updating the sufficient statistics. This paradigm
of ``looking'' at data point and performing an operation with them defines our computation cost in this problem. 
The total computational cost for the MLE is thus $2n$.\footnote{One might claim the cost is $2n+C$ where the $C$ represents the additional computation needed to complete the computations in Eq.~\eqref{eq:mle}. We omit this since this additional time is unchanged for all estimates and algorithms in this section. }
The risk associated with this estimate is \[ \Ex\left[(\hat{\mu}_{MLE}-\mu)^2\right]+\Ex\left[(\hat{\sigma}^2_{MLE}- \sigma^2)^2 \right] = \frac{\sigma^2}{n} + \frac{2\sigma^4}{n}.\]

% Throughout, 
% long term storage is treated as fixed for keeping track of the
% two statistics $\bar X$ and $\bar{X^2}$ (where the averaging
% is possibly done over different samples).

\subsection{Streaming setting}\label{subs:streaming}
We now consider what we call the streaming setting, a near zero-resource setting where local storage is extremely limited, allowing us access
to each data point exactly once.
This means that we can't compute the full MLE which requires two looks at each point to compute the first and second moments. 
We consider estimators with index set $S=\{1,\dotsc,n-1\}$ where for each $s\in S$ the estimate $\hat{\theta}_s$ is computed by first computing the moment sums
\[
    M_{1,s} = \sum_{i=1}^s X_i,\quad M_{2,s} = \sum_{i=s+1}^n X_i^2,%\quad\text{and}\quad n_1=\sum_{i=1}^n B_i.
\]
where $n$ is the total sample size.
We could have allowed the sums to be non-sequential but provided $s$ samples are used for $M_{1,s}$ and $n-s$ for $M_{2,s}$ the estimates are the same in risk and computation time due to the fact that the samples are exchangeable.
We assume that updating either $M_{1,s}$ or $M_{2,s}$ has the same cost so that the cost of computing these statistics is exactly $n$ as each data point is accessed only once
and again we can simply keep track of two sufficient statistics.

\begin{remark}\label{rem:cost}
We remark here that assigning a cost of $n$ to each of these estimators is, of course, an abstraction but a useful one nonetheless.
In reality the time to compute an estimate will be some function of $n$, $s$, and will also depend on the implementation in high-level and low-level languages down to the structure of the hardware being used. 
However, for a large range of $n$ and $s$ we believe it is reasonable to assume that the cost to compute $M_{1,s}$ and $M_{2,s}$ will be linear in $s$ and $n-s$ respectively. 
Boiling this down to the assumption that the cost is $n$ makes the following analysis quite clear but generalizing slightly does not change the overall flavor of the result as we discuss at the end of Section~\ref{sec:ef}.
As we go forward, this paper we will make similar assumptions about the computational cost that allow for numerical analysis but do not impact the overall framework.
\end{remark}

From these statistics we can define the unbiased streaming estimates for the mean and variance as as \begin{align*}
\hat\mu_s&=\frac{1}{s} M_{1,s}, \quad\text{ and}\quad\\
\hat\sigma^2_s&=\frac{s}{(s-1)(n-s)}
\left(M_{2,s}- \frac{n-s}{s} M_{1,s}^2\right).
\end{align*}
We note that unlike in the
example of the MLE where estimates of the mean and the variance are independent,
for the streaming setting the estimates of the mean and the second non-central moment
are independent while the estimates of the mean and the variance
are not. This will be explored in greater detail in Section \ref{sec:ef}
when discussing estimates of
 natural versus mean value parameters for general exponential families.
 
The risks of the mean and variance, both under quadratic loss, in the streaming setting 
 are given by
 \begin{align*}
&\R(\hat\mu_s,\mu)=\frac{\sigma^2}{s}\\
 &\R(\hat\sigma^2_s,\sigma^2)=\frac{2 s n\mu^2\sigma^2+2((s-1)s+n)\sigma^4}{(s-1)^2(n-s)}.
 \end{align*} 
 % % We let the combined risk
 % % of the two estimators be the sum 
 % % $\R(\hat\mu_s,\hat\sigma^2_s,\mu,\sigma^2)=\R(\hat\mu_s,\mu)+\R(\hat\sigma^2_s,\sigma^2)$.
 % The total risk can be defined as any convex function
 % of the two separate risks and encodes the relative
 % importance of estimating the mean and the variance
 % correctly. 

 Alternatively, we could consider the maximum likelihood estimate given the observed statistics $M_{1,s},M_{2,s}$ and $t$:
 \begin{equation}
          \hat{\mu}_{MLE,s} = \frac{1}{s} M_{1,s}\text{ and }\hat{\sigma}^2_{MLE,s} = \frac{1}{n-s}M_{2,s}-  \left(\frac{1}{t} M_{1,s}\right)^2.\label{eq:str_mle}
 \end{equation}
  As $n$ gets larges and $t/n$ tends to a constant $p\in(0,1)$, these estimates are essentially equivalent and will both have the same asymptotic risk.
 Indeed, asymptotically we have that the risk is approximately \[
     \frac{\sigma ^2 \left(4 \mu ^2+2 p \sigma ^2-p+1\right)}{np(1-p)}
 \]
 and one can then verify that in order to minimize this asymptotic risk as a function of $(\mu,\sigma)$ one should select $p=\frac{\sqrt{4 \mu ^2+1}}{\sqrt{4 \mu ^2+2 \sigma ^2}+\sqrt{4 \mu ^2+1}}$.
 Intuitively, this says that if the signal to noise ratio is very small, than $p$ is close to zero with most effort put on computing the second moment $M_{2,s}$.
 On the other hand if $\sigma^2\ll \mu^2$ and $\mu\gg0$ then $p$ will be close $1/2$.
 Finally, if  $\sigma^2\ll \mu^2$ and $\mu\approx 0$ then $p$ will be close to $0$.

\subsection{Risk/computation frontier}
% We can construct a risk-computational frontier
% for a slight generalization of the estimators described above.
The streaming setting is a special case of a slightly more general collection of estimators. 
In particular, we allow up to two looks and operations for each point with some points possibly only used for the computation of one of the two statistic, others may be used for both, while still other samples may be ignored completely.
The collection of estimates is indexed by pairs of sets $(S_1,S_2)$  where each $S_i$ is a non-empty subset of $\{1,\dotsc,n\}$ so the index set is  
\begin{equation}
    S=\left\{(S_1,S_2): S_1\subset[n], S_2\subset[n], |S_1|>0, |S_2|>0\right\}. \label{eq:index}
\end{equation}
For $s=(S_1,S_2)\in S$ we first compute
    \begin{align}
    M_{1,s} &= \sum_{i\in S_1}  X_i,\quad   M_{2,s}= \sum_{i\in S_2} X_i^2, \label{eq:normstat}
    \end{align}
and we let $n_1=|S_1\setminus S_2|$, $n_2=|S_2\setminus S_1|$ and $n_{12}=|S_1\cap S_2|$.
% where $B_{1,i}$ indicates whether $X_i$ was used for $M_{1}$ and $B_{2,i}$ idicates whether $X_i$ was used for $M_{2,s}$.
As in the previous setting, the particular sets $S_1$ and $S_2$ only impact the risk and computation time of the estimates in terms of $n_1$, $n_2$ and $n_{12}$.
The computational cost we assign to this procedure is \[ C= n_1+n_2+2n_{12},\] since $n_{12}$ indicates the number of samples used for both statistics while $n_1$ and $n_2$ indicate the number of samples used only for computing $M_{1,s}$ and $M_{2,s}$, respectively. 
Our maximum-likelihood-like estimates are then \[
     \hat{\mu}_s = \frac{1}{n_1+n_{12}} M_{1,s} \text{ and }\hat{\sigma}^2_s = \frac{1}{n_2+n_{12}} M_{2,s}- \hat{\mu}_s^2 
\]

\begin{table}[b!]\label{tab:mixedrisk}
\centering\begin{tabular}{cc|ccc}
 $\SNR$ & $\sigma^2$ & $n_1/n$ &$n_2/n$& $n_{12}/n$ \\ \hline
 $=0$ & $>1/2$ & \multicolumn{3}{c}{2 Change points}\\
 $=0$ & $=1/2$ & \multicolumn{3}{c}{Non-unique Solution}\\
 $=0$ & $<1/2$ & \multicolumn{3}{c}{2 Change points}\\ \hline
$\geq 1$ & $>0$  &0&0&$C/2$ \\\hline
 $\in(0,1)$ & $>1/2$ & 0& $a_C $&$(C-a_C)/2$\\
$\in(0,1)$ & $=1/2$ & 0 &0& $C/2$ \\
$\in(0,1)$ & $<1/2$ &$b_C$ &0&$(C-b_C)/2$ \\
 \end{tabular}
  \caption{The split of the data points between the three
 possible usages (mean only, non central second moment only, or both)
 is provided in columns 3 through 5. Both $a_C$ and $b_C$
 approach to 0 as $C \rightarrow 2n$. See Figure~\ref{fig:tradeoff} for a detailed look at particular parameter values in these different regimes.}
\end{table}

\begin{figure}[t!]
    \begin{center}
        (a) $\mu=1/10$ and $\sigma=1/2$
        \includegraphics[width=.5\textwidth]{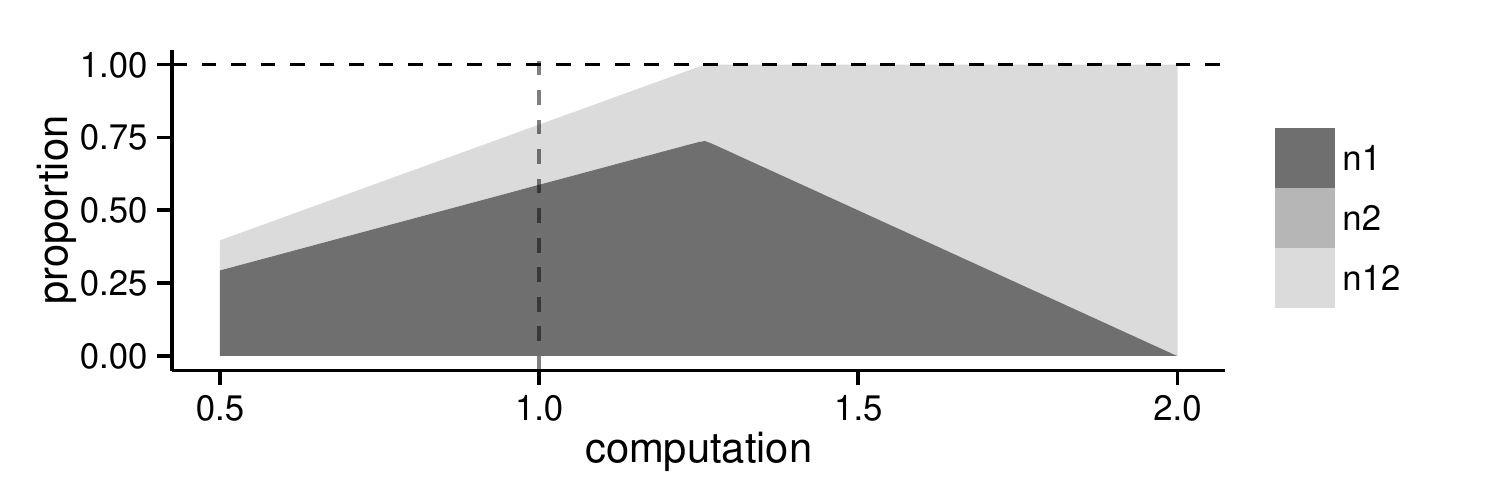}\includegraphics[width=.48\textwidth]{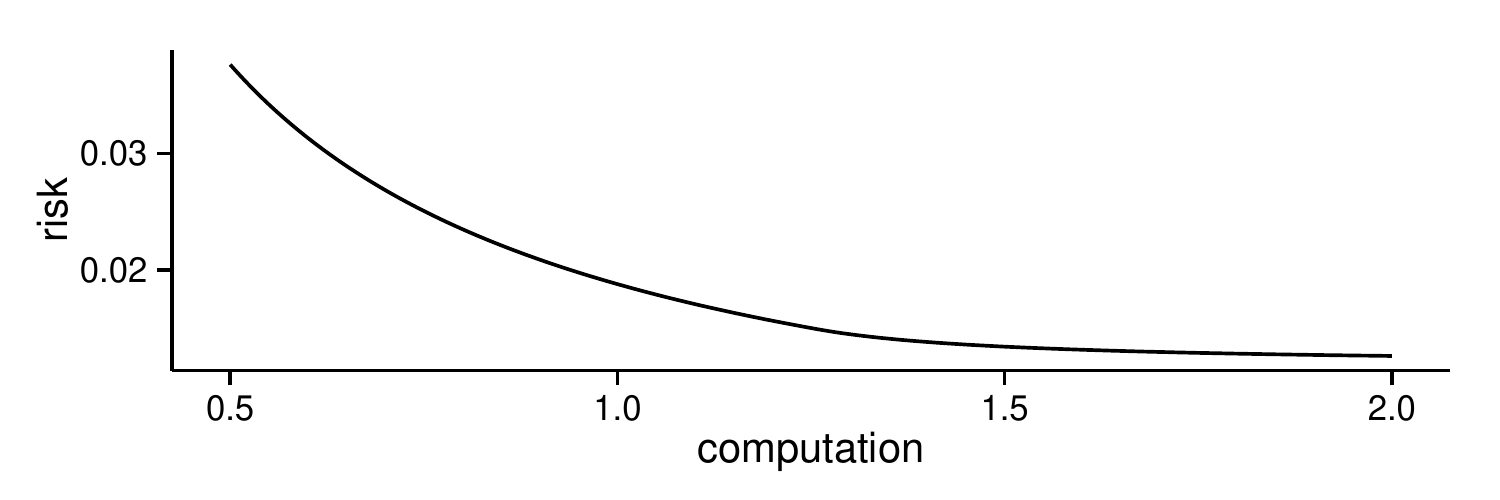}\\
        (b) $\mu=1/2$ and $\sigma=1/4$
        \includegraphics[width=.5\textwidth]{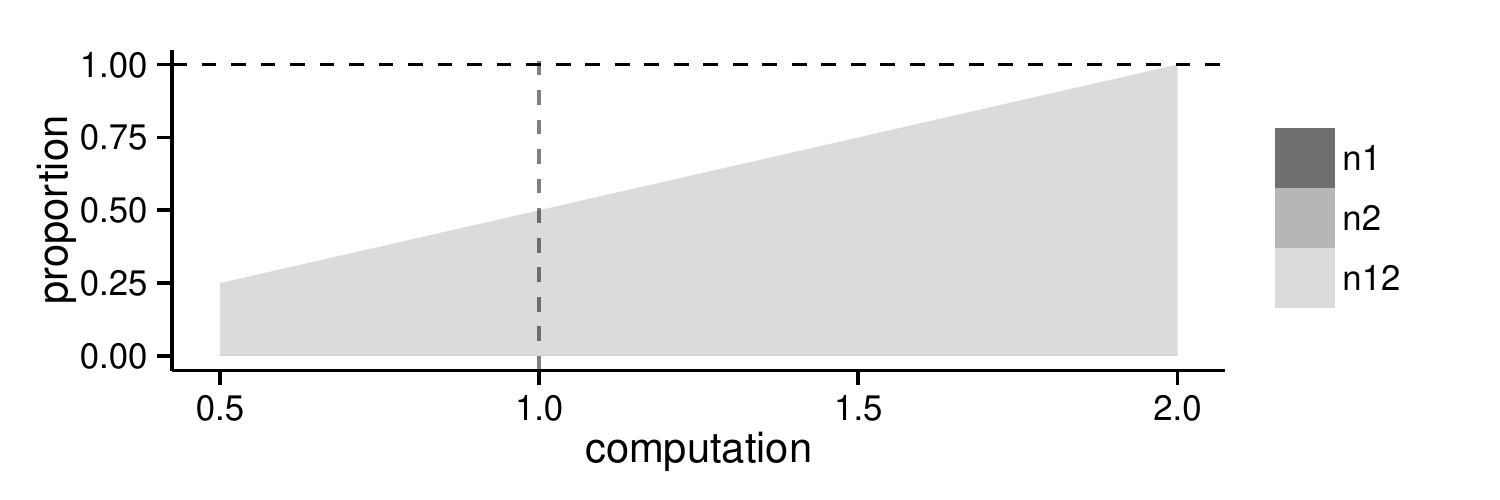}\includegraphics[width=.48\textwidth]{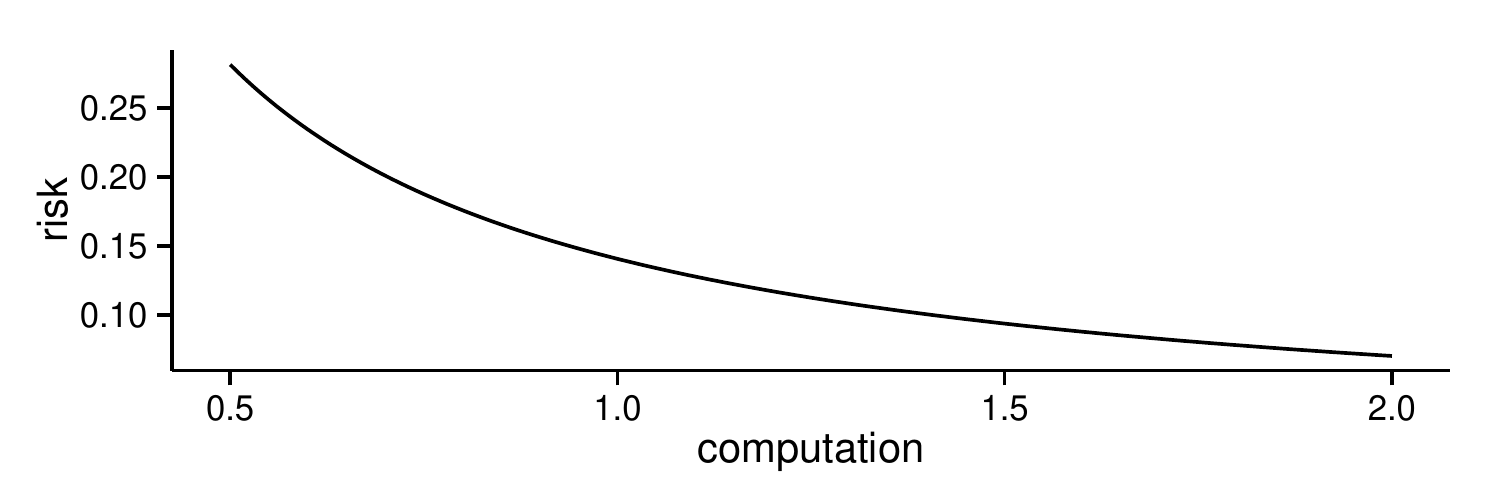}\\
        (c) $\mu=1/2$ and $\sigma=10$
        \includegraphics[width=.5\textwidth]{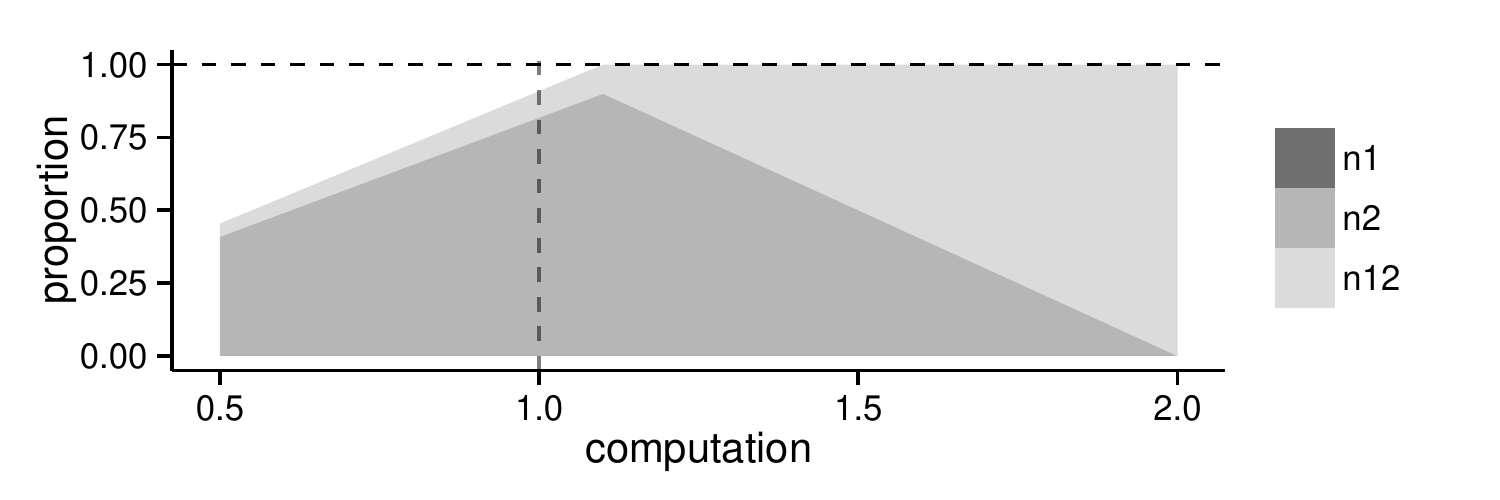}\includegraphics[width=.48\textwidth]{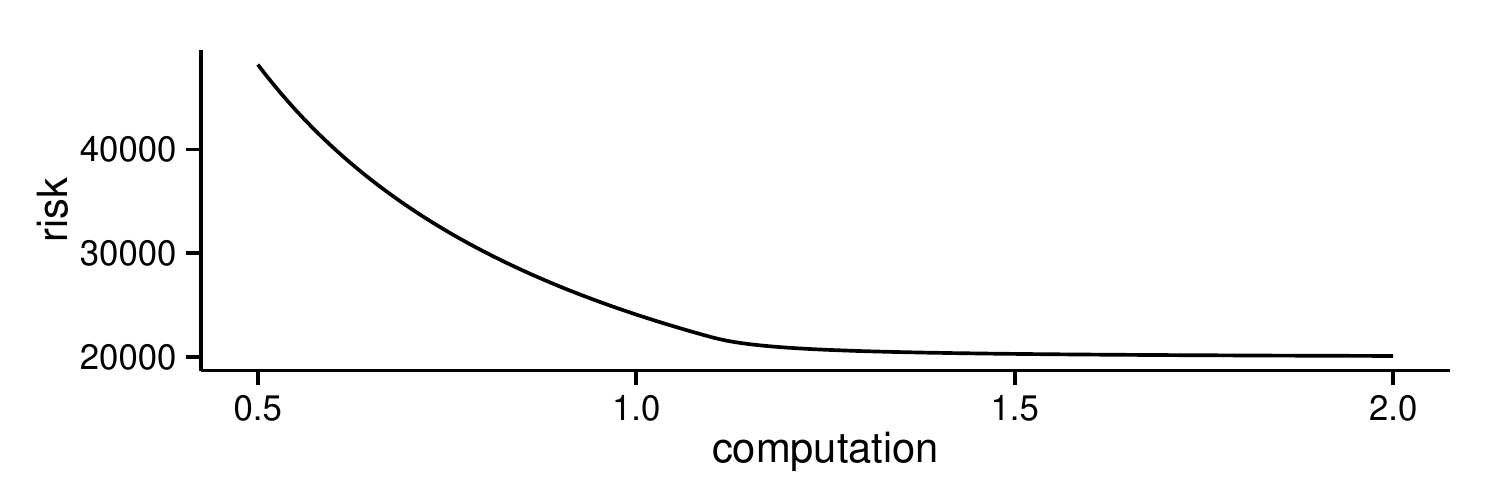}\\
        (d) $\mu=0$ and $\sigma=1$
        \includegraphics[width=.5\textwidth]{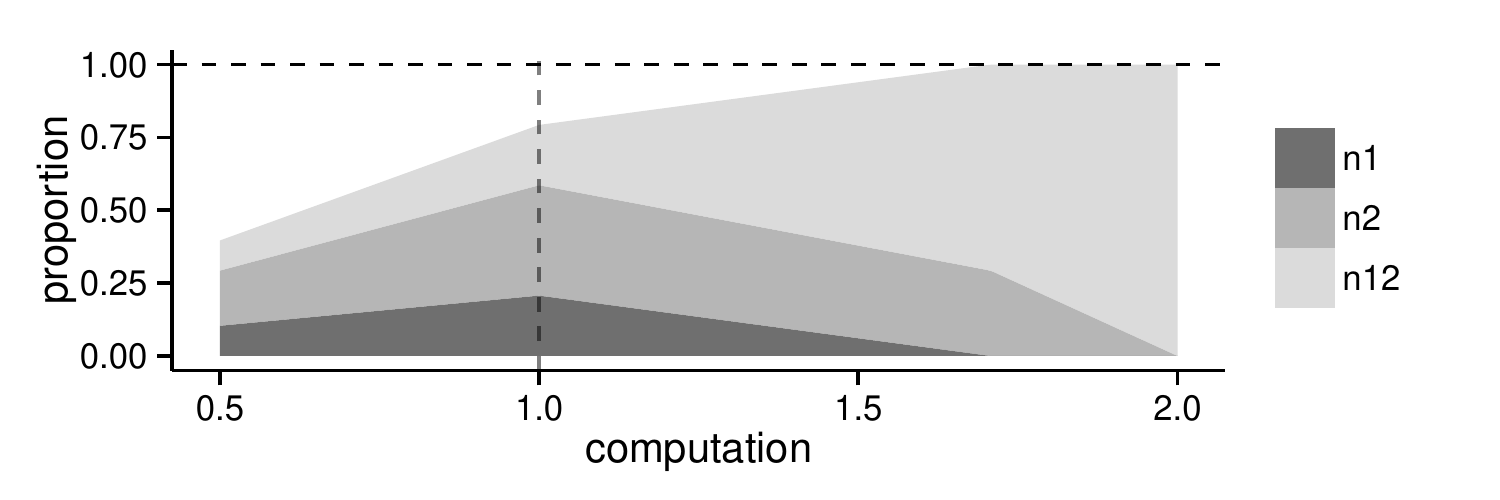}\includegraphics[width=.48\textwidth]{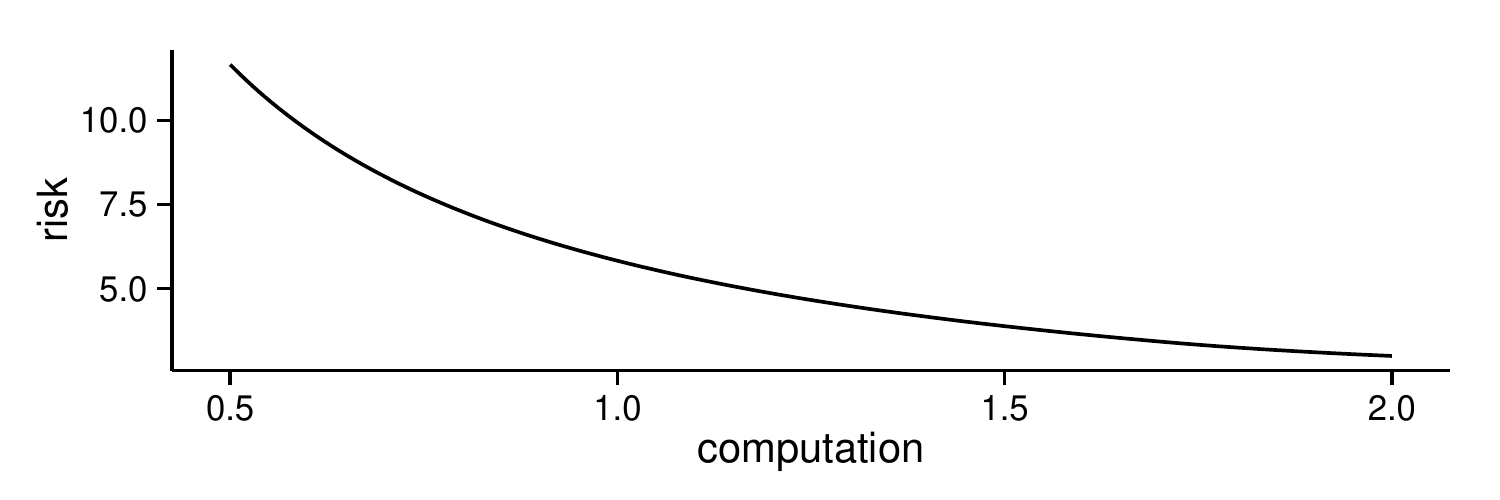}
    \end{center}
    \caption{In each row, the left panel illustrates the optimal proportions $n_1/n$ in red, $n_2/n$ in green and $n_{12}/n$ in blue as a function of the computational constraints given along the horizontal axis.
    The computational constraint is given as a proportion of the number of samples $n$, so that on the horizontal axis $1$ indicates a cost of $n$ and $2$ indicates a cost of $2n$, where the standard estimate is always optimal.
    The right panel illustrates the risk using the given estimate with the given proportions.
    Each of the rows corresponds to a different row in Table~\ref{tab:mixedrisk}.}
    \label{fig:normTradeoff}
\end{figure}

To get the risk of these estimators
we provide a glimpse of the general Fisher information 
result from the following section.
\begin{proposition}
Let $X_1,\dots,X_n\sim{\rm normal}(\mu,\sigma^2)$. For $s=(S_1,S_2)\in S$ from Eq.~\eqref{eq:index}, let $M_{1,s}$, $M_{2,s}$, $n_1$, $n_2$ and $n_{12}$ be as in Eq.~\eqref{eq:normstat}.
% We have that $n_1+n_{12}+n_2\leq n$ and let the computational
% constraint be as before $n_1+2n_{12}+n_2=C$.
For $n$ large, the covariance for estimates $(\hat{\mu}_s, \hat{\sigma}^2_s)$ of $(\mu,\sigma)$ is approximately
\begin{align} \label{eq:normInfo}
\left(
\begin{array}{cc}
 \frac{\sigma ^2}{n_1+n_{12}} & -\frac{2 n_2 \mu  \sigma ^2}{(n_1+n_{12}) (n_{12}+n_{2})} \\
 -\frac{2 n_2 \mu  \sigma ^2}{(n_1+n_{12}) (n_{12}+n_{2})}& 
 \frac{2 (n_{1}+n_{12}) \sigma ^4+4 (n_{1}+n_{2}) \mu ^2 \sigma ^2}{(n_{1}+n_{12}) (n_{12}+n_{2})} \\
\end{array}
\right)
\end{align}
and so the overall risk of the estimates is approximately \[
    \frac{\sigma ^2}{n_1+n_{12}}+ \frac{2 (n_{1}+n_{12}) \sigma ^4+4 (n_{1}+n_{2}) \mu ^2 \sigma ^2}{(n_{1}+n_{12}) (n_{12}+n_{2})} .
\]
%As $n\to \infty$, the maximum likelihood estimates for $(\mu,\sigma)$ based on $M_{1,s}$ and $M_{2,s}$ will asymptotically achieve the lower bound.
\end{proposition}
% This proposition is an application of Theorem~\ref{thm:expFamGen} and a change of parameters.
% Letting the trace of the matrix $I^{-1}$ be the asymptotic total risk (again we note that
% any convex combination of the diagonal matrices
% is appropriate as a measure of total risk and merely determines
% the relative importance of the mean and variance estimation) we can 

For a given $\mu$ and $\sigma$ we can compute $n_1$, $n_2$ and $n_{12}$ that minimize the total risk for any given computational cost $C$ satisfying $2\leq C\leq 2n$.
Several scenario, in terms of the signal
to noise ratio $\SNR=\mu/\sigma$ and variance $\sigma$ are presented
in Table~\ref{tab:mixedrisk}. In particular, we note that if $\SNR=1/\sqrt{2}$ we always choose to use less of 
the data and essentially construct an MLE.
Otherwise, there are different optimal $n_1$, $n_2$ and $n_{12}$ based on whether $\SNR$ is zero or between 0 and 1 and on whether $\sigma^2$ is greater than, equal to, or less than $1/2.$
To further illustrate the different regimes, Figure~\ref{fig:normTradeoff} shows the optimal portions of the samples used and the associated risk as the computational constraint varies.
In the next section we explore the tradeoff of risk and compute time for the more general setting of exponential families.

%% file: asymptot.tex
\section{Exponential family}
\label{sec:ef}

A $p$ parameter exponential family is a family of distributions $\{f_\theta\}_{\theta\in\Theta}$ on a space $\mathcal{X}$ each with a density with respect to an appropriate carrying measure $\mu$ which can be written in the form \[
    f_\theta(x) = h(x) \exp\{\theta^T t(x) - \Psi(\theta)\}
\]
where $h$ is a function from $\mathcal{X}$ to $\Re^+$, $\theta\in \Theta\subset \Re^p$ is the {\em natural parameter} for the model, $t=(t_1,t_2,\dotsc, t_p)^t$ is a real-valued $p$-dimensional sufficient statistic, ie. $t_k:\mathcal{X}\mapsto \Re$, and $\Psi(\theta)=\log\int_{\mathcal{X}} h(x) \exp\{\theta^T T(x)\}dx$, \citep{bickeldoksum}.

For a random sample $X_1,\dotsc, X_n \stackrel{iid}{\sim}f_\theta$, the statistic $T= \sum_{i=1}^n t(X_i)\in \Re^p$ is sufficient and
the maximum likelihood estimate $\theta$ is found by solving for $\theta$ in the equation $T/n=\Ex_{\theta}[t(X)]$.
As in the normal example, for our purposes it is convenient to reparameterize the model using the {\em mean value parameterization} with parameters $\tau=\tau(\theta)=\Ex_\theta[T(X)]\in \Re^{p}$ where the MLE for $\tau$ is simply $T/n$.
Note that the map $\tau:\theta\to \tau(\theta)$ is invertible so that $\tau$ is a proper reparametrization of the model.
% Optimization Problem Original
 
Using the same ideas as in our normal example, we can construct alternative statistics by computing each component of the sufficient statistic %, $T_1 ,T_2,\dotsc, T_p$,
on a subset of the full data set.
Formally, our estimates will be indexed by subsets $S_1,\dotsc,S_p \subset [n]$ so the index set is $\mathcal{S}=(2^{[n]})^p$.\footnote{Again, the actual sets are not critical but only their cardinality and the cardinality of the pairwise interections.}
For $S=(S_1,\dotsc,S_p)\in \mathcal{S}$, we define the statistics \[
  T_{S} = \left( \sum_{i\in S_1} t_1(X_i),%\sum_{i\in S_2} t_2(X_i),
  \dotsc,\sum_{i\in S_p} t_p(X_i) \right)
\] and the estimate for $\tau_k$ is $\hat{\tau}_{S,k}=|S_k|^{-1} T_{S,k}$. 
The estimate for $\theta$ is defined analagously, $\hat{\theta}_{S}= \tau^{-1}(\hat{\tau}_{S})$.

The covariance for $\hat{\tau}_{S}$ can be written in terms of the covarariance for $T(X)$ which is $I^{-1}(\tau)$, the inverse of the Fisher information matrix for $\tau$, and the cardinality of the sets $S_1,\dotsc, S_p$ and their pairwise intersections.
Specifically, the variance terms are $\mathrm{Var}(|S_k|^{-1} T_{S,k})= \frac{I^{-1}(\tau)_{kk}}{|S_k|^2}$ and the covariance terms are $\mathrm{Cov}(|S_k|^{-1} T_{S,k},|S_l|^{-1}  T_{S,l}) = \frac{|S_k \cap S_l| I^{-1}(\tau)_{kl}}{|S_k| \cdot |S_l|}$.

Note that the analog to the streaming case for a $p$-parameter exponential family is where $S_k\cap S_l=\emptyset$ for all $k\neq l\in[p]$ and in this case the covariance matrix for $\hat{\tau}_{S}$ will be diagonal as isevident by the fact that each statistic is computed on an independent sample.
However the covariance for $\hat{\theta}_S$ will usually not be diagonal, as can be verified in the case of the normal example.

For a given cost level $c$ and estimate $\theta=\theta(\tau)=\tau^{-1}(\tau)$ we can find the best subsets by solving an appropriate optimization problem. 
Frequently the compute times for each statistic will be different and so we can define the computational cost for each statistic in terms of the cost to compute $t_k(x)$ and the set sizes $|S_k|$ for $k\in[p]$.
Specifically, we denote by $c_k$ the runtime to perform the operation that computes $t_k(x_i)$ and adds it to the partial sum.
The risk for estimating $\theta$ is $R(\theta, \hat{\theta}_{S}) = \mathrm{tr}\left( \dot{\theta}(\tau)  \mathrm{Cov}(\hat{\tau}_S)\dot{\theta}(\tau)^T \right)$ but if we want to estimate another parameter $\eta=\eta(\tau)$ then the risk is given by $\mathrm{tr}(\dot{\eta}(\tau) \mathrm{Cov}(\hat{\tau}_S)\dot{\eta}(\tau)^T)$ where $\dot{\eta}(\tau)$ is the gradient of $\eta$ with respect to $\tau$.
Finally, some parameters may be more important to estimate than others and so we allow for the scaling of the covariance by a non-negative diagonal matrix $Q$ which indicates the relative importance of the different components of the parameter.
Together this yields the optimization problem
\begin{align}
\min_{S\in \mathcal{S}} \quad& R(\eta, \hat{\eta}_{S}) = \mathrm{tr}(Q \dot{\eta}(\tau)\Sigma \dot{\eta}(\tau)^T)\\
\text{ such that} \quad& \sum_{k=1}^p c_k|S_k| \leq c,\label{eq:constraint}\\
\mathrm{where}\quad % & S_1,\dotsc, S_p \subset [n] \quad \text{and}\\
 & \Sigma_{kl} = \begin{cases}
     \dfrac{I^{-1}(\tau)_{kk}}{|S_k|^2}, &\text{ if } k=l\\
         \dfrac{|S_k \cap S_l| I^{-1}(\tau)_{kl}}{|S_k| \cdot |S_l|} & \text{ otherwise.}
\end{cases}
\end{align}

In Section~\ref{sec:norm} we were able to solve this problem in the case of the normal distribution and estimation of the mean and variance parameters. 
Solving this problem for certain other distributions such as the multivariate normal is also relatively straightforward. 
In general, computing the Fisher information matrix and its inverse for the mean value parameterization of an exponential family is a nontrivial task. 
Additionally, the functions $\eta(\tau)$ are generally difficult to compute especially in high dimensions.
For example, \citet{montanari2014computational} shows that for certain classes of graphical models finding computing the map from the mean-value to the natural parameter space is an NP-hard problem in the dimension of the parameter space.
Nonetheless, the formulation of this optimization problem offers another step towards an understading the frontier of the risk-runtime tradeoff.
In the next two sections we will deviate slightly and consider estimation procedures with slightly less well understood 
% Optimization Problem Changed
% If we seek to estimate the parameters $\Eta$
% \section{Bivariate Normal as Correlation Changes}

%% file: hl_example.tex
% !TEX root = main.tex

% Hodges-Lehmann Example

\section{Hodges-Lehmann estimator}
\label{sec:hl}
As another investigation into possible tradeoffs between computation time and statistical risk we consider the Hodges-Lehmann (HL) estimate for the mean of a distribution. 
For a sample of size $n$, this estimates is defined as
\begin{equation}
	\hat{\theta}_{HL} = \mathrm{median} \left\{\frac{X_i+X_j}{2}: i,j\in[n] \right\}. \label{eq:hl_Est}
\end{equation}
This estimate is known to be very robust and often outperforms both the mean and the median in terms of statistical risk for data arising from distributions with contamination.
In this example we consider the contaminated distribution \[
	(1-\alpha) \mathcal{N}(0,1) + \alpha\mathcal{T}_3(4)
\] 
where $\mathcal{T}_3(4)$ is a central $t$ distribution with three degrees of freedom and then scaled by a factor of four.
Each mixture component has mean zero however approximately ten percent of any sample will be from a contaminated distribution with much heavier tails and higher variance. The proportion of data arising from the $t$ distribution is the contamination level $\alpha$.

\begin{figure}[tb]
	\centering
	\includegraphics[width=\textwidth]{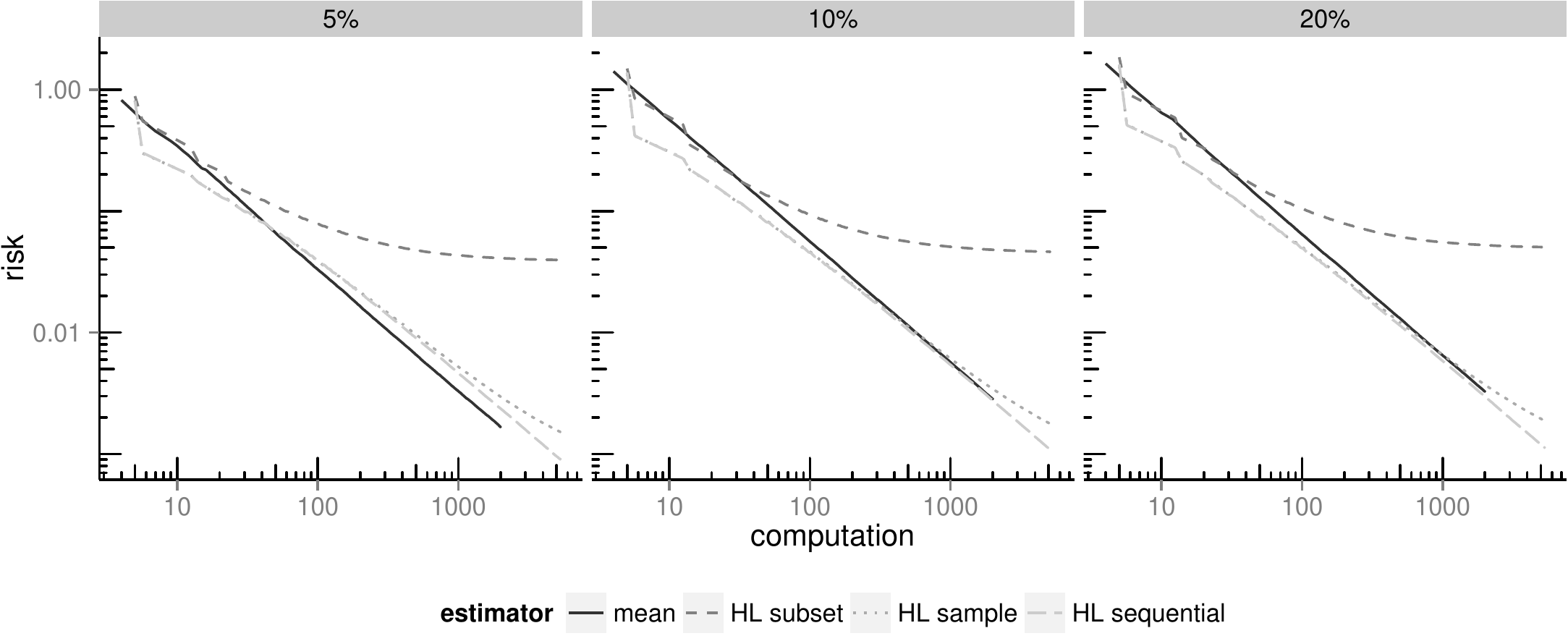}
	\caption{Computation and risk on the log-log scale for the sample mean and variations on the Hodges-Lehmann estimators. The computation is determined by the number of operations to compute the pairwise means and the number of comparisons to compute the median. The three panels correspond to different levels of contamination of 5\%, 10\% and 20\%. At very low costs and very high costs the HL sequential estimator is best but at moder costs, between 100 and 1000, the level of contamination can impact whether the HL sequential estimator or the mean estimator is to be preferred. }
	\label{fig:hl}
\end{figure}

The computation cost of HL estimator can be decomposed into two parts, the time to compute the pairwise sums, which requires $n(n+1)$ looks at the data, and the time to compute the median. 
The time to compute the median is dominated by the number of comparisons needs and will in practice depend on the data.
For the simulations below we add the the expected number of comparisons, as determined in \citet{knuth1972math} for the QuickSelect algorithm, to the computation cost.
Asymptotically, the expected time to compute the median of $n$ samples is approximately $3.38n$,
while in comparison, the mean requires only $n$ operations as described above. 
We considered a variety of estimates in order to reduce the computation time of the HL estimate and compare them in Figure~\ref{fig:hl}:
\begin{description}\setlength\itemsep{0em}
	\item[\em subset] We first select a subset of the data of $m$ and then sample without replacement $c/2$ from all $\binom{m}{2}$ pairs in this subset. 
	\item[\em sample] We sample with replacement $c/2$ pairs from all $\binom{n}{2}$ possible pairs from the entire data set.
	\item[\em sequential] At cost $c$ we use the $c/2$ pairs $(X_1,X_2), (X_3,X_4), \dotsc,(X_{c-1},X_{c})$
\end{description}
For each of these estimates we compute the mean for each of the selected pairs and then compute the median for that set.

We used a sample size of $n=2000$ and for the subset HL estimate we used $m=\lfloor \sqrt{2000} \rfloor$. For the mean we simply considered the sample mean of the first $c$ points for $c\in [n]$ and for the three HL estimates we used even costs from $2$ to $n$.
We simulated $5\times10^4$ replicates for each contamination level to estimate the risk associated with each estimator at each cost. 
Overall, the best estimates were either the sample mean or the sequential HL, at least up to the feasible costs for those methods. Choosing between the sample mean and the HL sequential depends on the cost restraints as well as the contamination level as shown in Figure~\ref{fig:hl}.
Overall, this example illustrates the intricacies of the computational-statistical trade-off frontier for even relatively straightforward settings.

%% file: matrix.tex
% !TEX root = main.tex

\section{Matrix inversion}
\label{sec:mat}
One of the most important linear algebra operations 
for statistical analysis is the matrix inverse. 
It is also frequently the bottleneck of statistical
procedures as the operation is naively of
complexity order $\bigO{n^3}$ (Gauss-Jordan elimination) and
optimally, if impractically, of order $\bigO{n^{2.373}}$ \citep{williams2012multiplying}.
All practical algorithms for matrix inversion
are based on iterative approaches. Each iteration can
naturally define a computational cost metric that allows
us to evaluate the statistical risk versus computational 
cost tradeoff within an algorithm. Since all iterative methods
are meant to converge to the same numerical value this also 
suggests a method for comparing across algorithms
when explicit costs cannot be defined. 

In this section we consider the problem of 
finding the least squares estimator in a standard
linear regression 
$$Y=X\beta+\epsilon$$ where the columns of $X$
are correlated. Each of the $p$ columns of $X$ represents an attribute
of an individual. The solution $\hat\beta=(X^tX)^{-1}X^tY$
is well known and requires the inversion of the Gram matrix
$S=X^tX$. We explore two iterative methods for matrix inversion. The
first is a naive Newton-Raphson (NR) algorithm that inverts the matrix $A$
via the iterative procedure $A^\inv_k=2 A^\inv_{k-1}-A^\inv_{k-1}AA^\inv_{k-1}$.
It is clear that by letting $k\rightarrow\infty$ we get
$A^\inv_k\rightarrow A^\inv$ such that $A^\inv A=I$ for $I$ the identity
matrix.

The second method builds on the power method for eigenvector
and eigenvalue approximation. It is well known that the 
eigenvector associated with the largest eigenvalue of a symmetric matrix $A$ 
can be computed via the iteration $v^{(1)}_{k+1}=Av^{(1)}_{k}/\|Av^{(1)}_{k}\|_2$
where $\|\cdot\|_2$ is the squared $L_2$ norm. To compute the
eigenvector associated with the second eigenvalue one first
computes $v^{(1)}_{\infty}$ and then performs the above iteration
replacing $A$ with $A-v^{(1)}_{\infty}v^{(1)t}_{\infty}$. A similar
expression is available for smaller eigenvalues. We consider
several stopping criteria for for this approach. First we consider
stopping the computation of the first eigenvector after $k$ steps,
then compute the second eigenvector based on $A-v^{(1)}_{k}v^{(1)t}_{k}$
also stopping after $k$, and so on. A second approach considers 
stopping the first iteration after $k$ steps, the second after $k-1$ steps, 
until the $p$th after $k-p$ steps. 
% A final approach considers stopping
% after $k,k-2,k-4,\dots,k-2p$. 
% The second method inverts the matrix by computing the 
% eigenvalues (and their associated eigenvalues) in descending order.
% The ``iterative'' portion of this algorithm refers to the fact that
% with additional computed eigenvalue/eigenvector pair we can construct
% a better approximation to the full inverse since for $S=U\Lambda U^t$,
% the unique
% eigen-decomposition of $S$, $S^\inv$ can be computed by $U\Lambda^\inv U^t$.
% Since $\Lambda$ is a diagonal matrix, the inversion is only as expensive
% as the computational cost of an eigenvalue/eigenvector pair.
% \subsection{Simulation study}

For the purposes of exposition we consider a matrix $X=ZD^{1/2}C^{1/2}$ where 
the entries $Z_{ij}\overset{iid}{\sim}\mathcal{N}(0,1)$,
$C=(1-\rho)I+11^t\rho$,
the compound symmetry correlation matrix,
and $D$ is a diagonal matrix with entries decreasing uniformly
from 4 to 2. As $\rho\nearrow 1$ the matrix
approaches rank deficiency which suggests that for larger values of $\rho$
algorithms that approximate the inverse via lower rank matrices are likely to
perform as well as full rank inversions. In this simulation we consider
$p=10$. 
The inversion via Newton-Raphson has on the order of $10^3$ steps, but
in practice no more than $20$ steps are needed for numeric conversion of the 
algorithm. The power method approaches have the same algorithmic complexity
but converge even faster in practice.
Throughout, the true value of $\beta$ is a uniformly
separated sequence from $-1$ to $1$ and we consider
the risk of estimating $\beta$ under quadratic loss. For independent
and identically distributed noise $\epsilon\sim \mathcal{N}(0,1)$ we know that the
risk of estimating $\beta$ is given by
$\sum_i{\rm diag}\left((X^tX)^\inv\right)_i$ and so we use this exact value 
to confirm that an inversion method has converged. 

\begin{figure}[htbp]
	\centering
	\includegraphics[width=.95\textwidth]{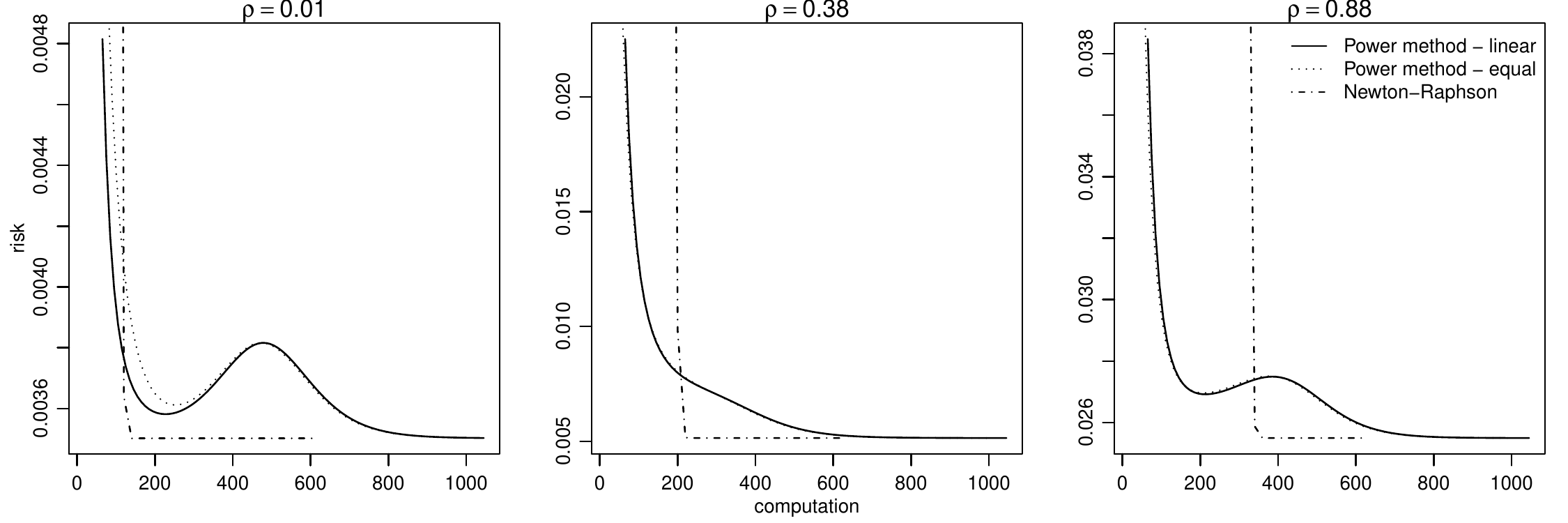}
	\caption{Three plots of simulation results
	comparing computational cost (as measured by the number of
	iterations of an algorithm) versus the risk under quadratic
	loss of estimating the coefficients in a linear regression.
	Each plot represents a different level of dependence among
	the columns of the design matrix $X$.}
	\label{fig:matinv}
\end{figure}

We simulate 10,000 datasets in order to estimate the risk for $\hat\beta$
across three values of $\rho$ between $0.01$ and $0.88$.
In Figure~\ref{fig:matinv} we see the outcomes of the experiment 
for both types of inversion procedures.
For the power method algorithm the computation cost is determined
by the total number of iterations of a single run. For example, the
linear method costs $\binom{k+p}{2}-\binom{k}{2}$ iterations, 
while the approach that stops every iteration after $k$ steps
costs $kp$. Each iteration of NR is assigned
a cost of $20$ since each one requires two matrix 
multiplications which involve 20 vector multiplications and each
iteration of the power method is a vector multiplication.
First it is evident that the risk is reduced non-linearly with
increases in computation costs for both methods. 
% What is of greater
% interest is that the two methods exhibit different computational 
% profiles across the values of $\rho$. 
The NR algorithm converges to the risk of the fully
inverted matrix faster than the two power-method approaches.
However it does so at the expense of poor performance
before full convergence. In particular, both NR and power-method
approaches are initialized naively. While NR has an initial risk (not
shown on the plots due to scale) hundreds of times greater than
the lowest risk, both power methods are within $30\%$ of 
the lowest risk after only a few iterations. Throughout the
plots we can see that NR is very dependent on the value of $\rho$
as that determines how well $S$ is approximated by a low rank 
matrix. The two power-methods appear agnostic to the value of $\rho$
with the exception of how smoothly they approach the lowest risk.
This can be explained by the interdependence between
iterations of the power method -- that is, a slight
improvement in the estimation of the first eigenvector (prior
to the convergence of the power method iterations)
does not guarantee an improvement in the estimation
of the second eigenvector.

%% file: conc.tex
\section{Conclusions}
This article proposed an interpretable framework for 
the tradeoff between computational cost and statistical 
risk. Our approach introduced exact computational cost
into the analysis of statistical methods. This is first
illustrated via the classical example of estimating the
mean and variance of a sample of normal random variables.
In this setting, we suggest that the use of a single data point
for the update of a sufficient statistic should incur a cost
of one. This allows us to compute exact and asymptotic risks
associated with mean and variance estimation under
a computational constraint. We extended this framework to 
general exponential families in Section~\ref{sec:ef}. 
We further illustrated our framework in the context of
robust estimators (Section~\ref{sec:hl}) and iterative
procedures (Section~\ref{sec:mat}).

We note, as we did in Remark~\ref{rem:cost}, that we have made simplifying
assumptions about the computational costs and runtimes of various procedures.
These assumptions allow for the subsequent analysis and we believe are
still helpful in guiding the choice of estimators. Sometimes a more detailed
and fine-grainded analysis may be desired that does not employ these abstractions.
In this case we believe that the practitioner could use benchmarking tools to 
precisely measure the runtimes of various aspects of their procedures which,
together with algorithmic analysis, can be used to employ our framework in
choosing the best procedure for the problem at hand.

Beyond the applications presented in this article our approach
can be employed whenever computational constraints are present.
In the context of experimental design, this framework can 
inform the number of observations or the number of subjects
needed for a study. For high throughput data it can assist
in deciding on a sampling mechanism when all data cannot be read 
into memory. The iterative procedures section suggests 
the development of analogues to standard methodology
(such as linear regression and spectral clustering) that
do not necessitate numerical convergence of intermediary
steps but that still preserve desirable statistical properties.

% Computer code for all of the results is available on GitHub at.
%%uncomment above once code is clean.

%% file: main_arxiv.bbl
\begin{thebibliography}{19}
\providecommand{\natexlab}[1]{#1}
\providecommand{\url}[1]{{#1}}
\providecommand{\urlprefix}{URL }
\expandafter\ifx\csname urlstyle\endcsname\relax
  \providecommand{\doi}[1]{DOI~\discretionary{}{}{}#1}\else
  \providecommand{\doi}{DOI~\discretionary{}{}{}\begingroup
  \urlstyle{rm}\Url}\fi
\providecommand{\eprint}[2][]{\url{#2}}

\bibitem[{Agarwal et~al(2014)Agarwal, Chapelle, Dud{\'{\i}}k, and
  Langford}]{alekh2014reliable}
Agarwal A, Chapelle O, Dud{\'{\i}}k M, Langford J (2014) A reliable effective
  terascale linear learning system. J Mach Learn Res 15:1111--1133

\bibitem[{Berthet and Rigollet(2013)}]{berthet2013computational}
Berthet Q, Rigollet P (2013) Computational lower bounds for sparse pca. arXiv
  preprint arXiv:13040828

\bibitem[{Bickel and Doksum(1976)}]{bickeldoksum}
Bickel PJ, Doksum KA (1976) Mathematical statistics. Holden-Day, Inc., San
  Francisco, Calif.-D\"usseldorf-Johannesburg, basic ideas and selected topics,
  Holden-Day Series in Probability and Statistics

\bibitem[{Bottou(2012)}]{leon2012large}
Bottou L (2012) Large-scale machine learning with stochastic gradient descent.
  In: Statistical learning and data science, Comput. Sci. Data Anal. Ser., CRC
  Press, Boca Raton, FL, pp 17--25

\bibitem[{Bresler et~al(2014)Bresler, Gamarnik, and Shah}]{bresler2014hardness}
Bresler G, Gamarnik D, Shah D (2014) Hardness of parameter estimation in
  graphical models. arXiv preprint arXiv:14093836

\bibitem[{Chandrasekaran and Jordan(2013)}]{chandra2013comp}
Chandrasekaran V, Jordan MI (2013) Computational and statistical tradeoffs via
  convex relaxation. Proc Natl Acad Sci USA 110(13):E1181--E1190,
  \doi{10.1073/pnas.1302293110},
  \urlprefix\url{http://dx.doi.org/10.1073/pnas.1302293110}

\bibitem[{Horev et~al(2015)Horev, Nadler, Arias-Castro, Galun, and
  Basri}]{inbal2015detection}
Horev I, Nadler B, Arias-Castro E, Galun M, Basri R (2015) Detection of long
  edges on a computational budget: a sublinear approach. SIAM J Imaging Sci
  8(1):458--483

\bibitem[{Kleiner et~al(2014)Kleiner, Talwalkar, Sarkar, and
  Jordan}]{kleiner2014scalable}
Kleiner A, Talwalkar A, Sarkar P, Jordan MI (2014) A scalable bootstrap for
  massive data. Journal of the Royal Statistical Society: Series B (Statistical
  Methodology)

\bibitem[{Knuth(1972)}]{knuth1972math}
Knuth DE (1972) Mathematical analysis of algorithms. In: Information processing
  71 ({P}roc. {IFIP} {C}ongress, {L}jubljana, 1971), {V}ol. 1: {F}oundations
  and systems, North-Holland, Amsterdam, pp 19--27

\bibitem[{Langford et~al(2009)Langford, Li, and Zhang}]{langford2009sparse}
Langford J, Li L, Zhang T (2009) Sparse online learning via truncated gradient.
  J Mach Learn Res 10:777--801

\bibitem[{Lehmann and Casella(1998)}]{lehmann1998tpe}
Lehmann EL, Casella G (1998) Theory of point estimation, 2nd edn. Springer
  Texts in Statistics, Springer-Verlag, New York

\bibitem[{Montanari(2014)}]{montanari2014computational}
Montanari A (2014) Computational implications of reducing data to sufficient
  statistics. arXiv preprint arXiv:14093821

\bibitem[{Scott et~al(2013)Scott, Blocker, Bonassi, Chipman, George, and
  McCulloch}]{scott2013bayes}
Scott SL, Blocker AW, Bonassi FV, Chipman HA, George EI, McCulloch RE (2013)
  Bayes and big data: The consensus monte carlo algorithm. In: EFaBBayes 250
  conference, vol~16

\bibitem[{Shender and Lafferty(2013)}]{shender2013computation}
Shender D, Lafferty J (2013) Computation-risk tradeoffs for
  covariance-thresholded regression. In: Proceedings of The 30th International
  Conference on Machine Learning, pp 756--764

\bibitem[{Toulis and Airoldi(2014)}]{toulis2014implicit}
Toulis P, Airoldi EM (2014) Implicit stochastic gradient descent for principled
  estimation with large datasets. arXiv preprint arXiv:14082923

\bibitem[{Wainwright and Jordan(2008)}]{wainwright2008graphical}
Wainwright MJ, Jordan MI (2008) Graphical models, exponential families, and
  variational inference. Foundations and Trends{\textregistered} in Machine
  Learning 1(1-2):1--305

\bibitem[{Wang et~al(2014)Wang, Berthet, and Samworth}]{wang2014statistical}
Wang T, Berthet Q, Samworth RJ (2014) Statistical and computational trade-offs
  in estimation of sparse principal components. arXiv preprint arXiv:14085369

\bibitem[{Williams(2012)}]{williams2012multiplying}
Williams VV (2012) Multiplying matrices faster than coppersmith-winograd. In:
  In Proc. 44th ACM Symposium on Theory of Computation, Citeseer

\bibitem[{Yang et~al(2015)Yang, Wainwright, and Jordan}]{yang2015computational}
Yang Y, Wainwright MJ, Jordan MI (2015) On the computational complexity of
  high-dimensional bayesian variable selection. arXiv preprint arXiv:150507925

\end{thebibliography}
